\documentclass{article}

\input{tcilatex}

\begin{document}

\textbf{ALGEBRAS OF CHARGES}

\bigskip

E. A. Novikov

\bigskip

Institute for Nonlinear Science, University of California - San Diego, La
Jolla, CA 92093 - 0402

\bigskip

Two different algebras are applied to the system of charges. Such
generalizations of classical theory eliminate infinite self-energy of
electrons. Paradoxical radiative self-acceleration of an electron is also
eliminated in this way. One of the algebras predicts some new effects of
high order.

\bigskip

\bigskip

Infinite self-energy of electrons is a long-standing problem in the
classical theory of electromagnetic field (EMF) [1]. The paradoxical effect
of radiative self-acceleration of an electron [2] has a similar nature. As a
remedy for these problems, two different algebras of charges are introduced
in this paper. One of these algebras predicts some new effects of high order.

Consider system of partial differential equations with real coefficients,
which we will call real system (RS). The Maxwell equations are example of
RS. Solution of RS, generally, can be a combination of real fields (RF) and
imaginary fields (IF). These two components can interact directly if RS is
nonlinear. The idea of IF came from the modeling of the effects of
consciousness on electric currents in human cortex [3] (collective effects
of billions of interconnected nonlinear neurons). For linear RS, the effect
of IF on RF can appear if a nonlinear operation is applied to a solution.
This is exactly the situation with problems of self-energy and
self-acceleration of electrons.

Firstly, consider electrostatic field. The Maxwell equations have the form:

\begin{equation}
div\mathbf{E}=4\pi \rho ,\text{ \ }curl\mathbf{E}=0  \tag{1}
\end{equation}%
where $\mathbf{E}$ is the electric field and $\rho $ is the charge density.
Hence, introducing scalar potential $\phi $, we get:

\begin{equation}
\mathbf{E}=-\nabla \phi ,\text{ \ \ }\Delta \phi =-4\pi \rho  \tag{2}
\end{equation}

For a system of point charges:

\begin{equation}
\phi =\sum_{a}\frac{e_{a}}{R_{a}}  \tag{3}
\end{equation}%
where $R_{a}$ is the distance from the charge $e_{a}$ to a point at which we
are determining $\phi $. For the energy of the system of charges we have:

\begin{equation}
U=\frac{1}{8\pi }\int E^{2}dV=\frac{1}{2}\int \rho \phi dV=\frac{1}{2}%
\sum_{a}e_{a}\phi _{a}  \tag{4}
\end{equation}%
where $\phi _{a}$ is the potential of the field produced by all the charges,
at the point where the charge $e_{a}$ is located. From (3) and (4) we see
that the self-energy of charge produces infinite contribution. This textbook
derivation shows how simple and profound is the problem. Note that energy
(4) is nonlinear operator. The traditional explanation for this problem is
that the theory is incorrect for small distances and infinite self-energy
must be ignored. Then, (3) and (4) give finite energy:

\begin{equation}
U=\sum_{a>b}\frac{e_{a}e_{b}}{R_{ab}}  \tag{5}
\end{equation}%
( $R_{ab}$ is the distance between charges). However, such approach seems
unsatisfactory. Similar problem with nonlinear operation arises in a more
hidden form.

The effect of radiative self-acceleration is more subtle and corresponding
derivation is based on use of the retarded potential. In the system of
reference moving with the charge at the given moment, the classical result
is [2]:

\begin{equation}
m\frac{d\mathbf{v}}{dt}=\frac{2e^{2}}{3c^{3}}\frac{d^{2}\mathbf{v}}{dt^{2}},
\tag{6}
\end{equation}%
where $m$ is the mass of the particle and $c$ is the velocity of light. This
equation, apart from trivial solution $\mathbf{v}=0$, has solution with
exponentially increasing self-acceleration. Note again, that this unphysical
effect is produced by nonlinear operation and corresponding force is
proportional to $e^{2}$. The traditional explanation of this paradox is that
(6) can be used only when the effect is small. However, in our opinion, (6)
can not be used at all.

To explain the last statement, note that (6) is obtained by expansion of the
retarded potential in terms of negative powers of $c$, namely, on the third
order of this expansion. But, the exact solution gives a different result.
Let us start with the Lienard-Wiechert potentials [1] for a charge moving
with velocity $\mathbf{v=}d\mathbf{r}_{o}/dt$:

\begin{equation}
\phi =\frac{ce}{cR-\mathbf{v\cdot R}},\;\mathbf{A}=\frac{e\mathbf{v}}{cR-%
\mathbf{v\cdot R}}  \tag{7}
\end{equation}%
Here $\mathbf{R}$ is the radius vector, taken from the point $\mathbf{r}_{o}$
where the charge is located to the point of observation, and all quantities
at the right sides of the equations must be evaluated at time $\tau $,
determined from equation:

\begin{equation}
c\tau +R(\tau )=ct  \tag{8}
\end{equation}

In the evaluation of electric and magnetic fields

\begin{equation}
\mathbf{E}=-\frac{1}{c}\frac{\partial \mathbf{A}}{\partial t}-grad\phi ,\;%
\mathbf{H}=curl\mathbf{A,}  \tag{9}
\end{equation}%
we use relation $\partial \tau /\partial t=cR/(cR-\mathbf{v\cdot R})$, which
follows from (8). Having in mind that potentials (7) satisfy the Lorentz
condition, the results can be simplified by using the frame of reference
moving with the charge at the given moment. Simple manipulation gives [4]:

\begin{equation}
\mathbf{E=}\frac{e\mathbf{R}}{R^{3}}+\frac{e\mathbf{R\times \{R\times }\frac{%
d\mathbf{v}}{d\tau }\}}{c^{2}R^{3}},  \tag{10}
\end{equation}

\begin{equation}
\mathbf{H}=\frac{1}{R}\mathbf{R\times E}  \tag{11}
\end{equation}

The first part of the electric field (10) corresponds to the electrostatics
(in the moving frame of reference). The second part in (10) represents the
effect of radiation. Components of this field we denote by $E_{i}^{(rad)}$
and rewrite in the form:

\begin{equation}
E_{i}^{(rad)}=\frac{e}{c^{2}R}(n_{i}n_{k}-\delta _{ik})\frac{dv_{k}(t-R/c)}{%
dt}  \tag{12}
\end{equation}

Here $n_{i}=R_{i}/R$, $\delta _{ik}$ is the unit tensor and summation is
assumed over the repeated subscripts from 1 to 3. We also used (8) in
transforming the argument of the velocity. For evaluation of the radiative
self-force we must take the limit of $eE_{i}^{(rad)}$ when $R\rightarrow 0$.
Isotropy of the space gives in the limit: $n_{i}n_{k}\approx \frac{1}{3}%
\delta _{ik}$. Expanding the acceleration term in (12), we get:

\begin{equation}
e\mathbf{E}^{(rad)}\approx -\frac{2e^{2}}{3c^{2}R}\frac{d\mathbf{v}}{dt}+%
\frac{2e^{2}}{3c^{3}}\frac{d^{2}\mathbf{v}}{dt^{2}}  \tag{13}
\end{equation}%
The second part in (13) reproduces (6). But, the first part is divergent and
has the same type of $e^{2}/R$ - divergency as the self-energy.

One way of eliminating such divergencies in linear RS is by introducing
appropriate IF. Particularly, let us make substitution:

\begin{equation}
e_{a}\rightarrow e_{a}(1+i_{a})  \tag{14}
\end{equation}%
where $i_{a}$ are imaginary units with condition:

\begin{equation}
i_{a}i_{b}=-\delta _{ab}  \tag{15}
\end{equation}%
In this paper we will be interested only in real components of nonlinear
operations. Possible interpretations of IF will be considered in future
(compare with Ref. 3). Clearly: $\func{Re}\{(1+i_{a})(1+i_{b})\}=1-\delta
_{ab}$. We see that (14) eliminates self-energy without changes in mutual
interactions (5). Radiative self-acceleration is also eliminated by (14).
Note that (14) is applicable only at locations of corresponding charges.

Condition (15) creates commutative nonassociative algebra. To show this, we
introduce complex numbers defined by formula $\alpha =\alpha
_{o}+\sum_{a}\alpha _{a}i_{a}$, where coefficients $\alpha _{a}(a=0,1,...)$
are real. From symmetry (15) follows commutativity condition: $\alpha \beta
=\beta \alpha $. Nonassociativity is clear from example: $%
(i_{1}i_{1})i_{2}=-i_{2},\;i_{1}(i_{1}i_{2})=0$. Simple calculation show
that: $(\alpha \alpha )(\beta \alpha )=((\alpha \alpha )\beta )\alpha $.
This equality insures that algebra, created by (15), is so called Jordan
algebra [5].

It is interesting to note that this algebra can predict some new effects of
high order. For example, in some nonlinear system we can get
self-interaction proportional to $e^{3}$. The algebra predicts that
corresponding term should be multiplied by $\func{Re}\{(1+i_{a})^{3}\}=-2$.
Generally, for high-order self-effects we have: $\func{Re}%
\{(1+i_{a})^{q}\}=2^{q/2}\cos (q\pi /4)$.

There is another way of dealing with indicated above unphysical effects,
namely, by introducing simple condition:

\begin{equation}
e_{a}e_{a}=0  \tag{16}
\end{equation}%
(no summation over subscript $a$). Clearly, condition (16), applicable at
location of the charge, eliminates self-energy and self-acceleration.
Assuming additionally that $e_{a}e_{b}=e_{b}e_{a}$, we get commutative
associative algebra, which, probably, has a name attached to it. Elements of
this algebra are:

\begin{equation}
\alpha =\alpha _{o}+\sum_{a}\alpha _{a}^{(1)}e_{a}+\sum_{a<b}\alpha
_{ab}^{(2)}e_{a}e_{b}+...+\alpha _{1...n}^{(n)}e_{1}...e_{n}  \tag{17}
\end{equation}%
where $\alpha _{...}^{(.)}$ are real coefficients and $n$ is the number of
charges. Formally, we do not use imaginary units in (16) and (17). However,
if $e_{a}\neq 0$ and $e_{a}e_{a}=0$, than $e_{a}$ is not real in ordinary
sense. So, we use the name IF in both considered cases.

Infinity of self-energy and paradoxical self-acceleration of an electron
tell us something about the nature of elementary particles and corresponding
fields. The presented above procedure of elimination of unphysical effects
(in terms of appropriate algebra) is only the first step. Next steps involve
problems in quantum field theories.

\bigskip

\bigskip

\textbf{REFERENCES}

\bigskip

[1] L. D. Landau and E. M. Lifshitz, The Classical Theory of Fields,
Pergamon Press, 1987.

[2] Ref.1, pages 204 - 208.

[3] E. A. Novikov, arXiv:nlin.PS/0309043; arXiv:nlin.PS/0311047;
arXiv:nli.PS/0403054; Chaos, Solitons \& Fractals, v. 25, p. 1-3 (2005);
arXiv:nlin.PS/0502028.

[4] Formula (10) corresponds to formula (63.7) in Ref. 1.

[5] Encyclopedic Dictionary of Mathematics, v. 2, article 231, Jordan
Algebras, MIT Press, 1987.

\end{document}